\documentstyle[multicol,aps,prb,epsf,graphics,psfig]{revtex}
\input epsf

\newcommand{\be}{\begin{equation}}
\newcommand{\bel}[1]{\begin{equation}\label{#1}}
\newcommand{\ee}{\end{equation}}
\newcommand{\bea}{\begin{eqnarray}}
\newcommand{\ba}{\begin{array}}
\newcommand{\eea}{\end{eqnarray}}
\newcommand{\ea}{\end{array}}

\begin{document}
\draft

\title{  Premium Forecasting of an Insurance Company: Automobile Insurance}
\author{M. Ebrahim Fouladvand and Amir H. Darooneh  }

\address{ Department of Physics, Zanjan University, P.O.
Box 45196-313, Zanjan, Iran.\\
URL: www.znu.ac.ir, e-mail: foolad@mail.znu.ac.ir}

\maketitle

\begin{abstract}

We present an analytical study of an insurance company. We model
the company's performance on a statistical basis and evaluate the
predicted annual income of the company in terms of insurance
parameters namely the premium, total number of the insured,
average loss claims etc. We restrict ourselves to a single
insurance class the so-called {\it automobile insurance }. We show
the existence a crossover premium $p_c$ below which the company is
loss-making. Above $p_c$, we also give detailed statistical
analysis of the company's financial status and obtain the
predicted profit along with the corresponding risk as well as ruin
probability in terms of premium. Furthermore we obtain the optimal
premium $p_{opt}$ which maximizes the company's profit.

\end{abstract}
 ]
\begin{multicols}{2}

\section{Introduction}
Rapidly expanding field of {\it interdisciplinary physics } has
witnessed the desire of statistical physicists to apply their
knowledge to other areas of analytical sciences which till
recently used to fall outside the traditional domain of physics. A
broad class of systems exhibiting complex dynamics, associated
with the presence of many interacting constituents, has shown to
be successfully investigated with the concepts and techniques of
statistical physics. Protein folding, financial markets
\cite{jean,stanely}, flow of vehicular traffic and granular
material \cite{andereas}, surface growth of interfaces are a few
examples of numerous applications of physical methods to the
paradigm of inter-disciplinary researches. In particular, economic
systems has remarkably spurred the interest of statistical physics
community. Specifically, stock markets has been the main attractor
of these interests and nowadays there is relatively a rich amount
of results, both analytical and numerical on modelling of
financial markets. Large number of traders with a diversity of
trading strategies, conflicting interests and heterogeneous
anticipations are generic aspects of financial market leading to
unpredictable price fluctuations and other out of equilibrium
phenomena which have challenged physicists as well as economists
for explanation. Nevertheless, stock markets constitute only a
part of economy and there are other major financial systems which
can be studied in the context of statistical physics. In this
paper, we present both analytical and numerical investigations of
one such economic systems: {\it an insurance market }. Human
beings are continuously confronted to different types of dangers
which threaten his/her life in many aspects. Illness, earthquake,
car accident, burglary, death and many others are simple examples
of inevitable dangers to which the human life is exposed. The
occurrence of these adverse events is random in nature and
frequently it is impossible to predict even the probability of
their occurrence. Insurance is way to provide guarantee of
compensation for the losses of these damages. The purpose of
insurance is to indemnify policy holders against the occurrence of
adverse events. There is tremendously variety in the events that
are covered by insurance. Common to all insurance systems, there
exists a {\it risk } i.e., a condition in which there is a
possibility of an adverse deviation from an expected outcome. Some
or all of the risk is transferred from the insured to the insurer
with an expectation that through the pooling of risks, the insurer
will improve the estimation of expected total losses. In this
paper we aim to model the performance of a virtual insurance
company in order to find a better insight into the stochastic
nature of the problem.

\section{ Formulation of the Model}

We now present our model which simulates the performance of an
insurance company. In this paper we restrict ourselves to a single
insurance class the so-called automobile insurance. In principle,
a number of clients (policy holders) insure their cars against car
accidents. The company issues the corresponding insurance policies
and receives an annual premium $p$ from the policy holders.
Generally speaking, the amount of premium depends on many
characteristics of the generic car such as type, model, production
year etc.
 However for the sake of simplicity, in this paper we restrict
ourselves to a category of cars for which the premium is almost a
constant amount. The company's performance is modelled by the
series of loss events which are randomly occurred in the course of
time. Each loss event i.e.; damages to cars due to accidents, is
incorporated with a loss amount which should be regarded as the
amount of loss confirmed by company for each accident. The loss
size is itself a random variable ranging from tiny to large
amounts. Upon occurring a loss event, the company pays the claim
amount to the policy holder which reduces the company's capital.
On the other hand, the premium income which is injected to the
company raises its capital. It is the competition of these two
factors which determines the financial status of the company. A
low frequency of loss events together with a high premium income
makes the company profitable while a low premium income and high
frequency of loss events gives rise to a ruin state. Evidently the
amount of premium plays a dominant role in insurance industry .
The task of premium calculations and the related strategies have
been challenging and controversial subjects for insurance
companies and can drastically affect the financial status of the
company \cite{otovianni,klugman,daykin}. In a competitive economy,
there are plenty of insurance companies each of which offer their
own premium. If a company increases the premium, the number of
clients willing to insure their cars decreases (due to competition
of the other companies offering lower premiums). In other words,
the company's attempt to raise its income through increasing the
premium may fail due to decrease of the number of insured which
obviously reduces the premium income. Therefore, increasing the
premium is a highly risky action and it is of prime importance for
the company's managers to have an estimation of the risk amount.
Thus it is indispensable for a company to have a quantified
knowledge on how the premium variations affect the long-term
profit. To be more specific, in each day, a random number of
clients insure their car in the company. We denote this number by
${\cal I}_t$ for $t$-th day. We assume that the number of daily
policy issued obeys a simple statistics with premium-dependent
characteristics such as mean, standard deviation etc. In this
paper we assume that ${\cal I}_t$ is uniformly drawn from the
interval $[a,b]$ with $b=\frac{3 {\bar {\cal I}}}{2}$ and
$a=\frac{{\bar {\cal I}}}{2}$ such that ${\bar {\cal I}}_t={\bar
{\cal I}}$ for all $t$. As a simple choice, we take an exponential
form ${\bar {\cal I}}(p)={\cal I}_0e^{-\frac{p-p_0}{p_0 \tau}}$
for the average number of daily policy holders. $p_0$ is an
arbitrary reference point, ${\cal I}_0$ is average number of
insured for $p=p_0$ and the parameter $\tau$ measures the
predicted fall of the number of policy holders upon increasing the
premium. For instance $\tau=0.45$ corresponds to a $20$ percent
fall upon a $10$ percent premium increase over $p_0$. We now
discuss on the distribution of the daily number of loss events in
$t-$th day which is denoted by ${\cal A}_t$. Evidently it
implicitly depends on the previous number of issued policies
$\sum_{t'<t} {\cal I}_{t'}$. However, for the sake of simplicity,
we make a simple assumption such that ${\bar {\cal A}}_t= \theta
{\bar {\cal I}}$ where constant $\theta $ measures the ratio of
loss events to number of issued policies at an average level. We
recall that $\theta$ can be interpreted as the probability of
occurrence of loss event (per car). We note that $\theta$ can be
estimated via empirical data of realistic insurance companies.
Analogous to ${\cal I}_t$, it is assumed that ${\cal A}_t$ is
uniformly distributed in the interval $[a\theta,b\theta]$ such
that ${\bar {\cal A}}_t=\theta{\bar {\cal I}}$ for all $t$. The
loss amount itself should be considered as a stochastic variable.
In this paper we adopt Erlang's model \cite{straub,arlang} for
statistical description of loss amounts. According to this model,
the distribution of loss amount obeys an exponential function of
the form $ e^{-\frac{x}{\xi}} $ where $\xi$ denotes the average
loss.


\section{ Statistical Description }

In this section we present the analytical results which give the
annual capital of the company in terms of insurance
characteristics i.e., premium, statistics of the number of
insured, statistics of loss events etc. For this purpose, in each
working day we randomly draw two integer number ${\cal I}$ and
${\cal A}$ as the numbers of issued policies and loss events.
Correspondingly, for each loss event, we attribute a random loss
amount (drawn from Erlang distribution function). The sum of loss
amounts gives us the daily loss which is denoted by ${\cal L }$.
Denoting the number of annual working days by ${\cal N }$ and the
capital at the end of $t$-th day by ${\cal C }_t$, we simply have
 \be
{\cal C}_t={\cal C}_{t-1} + p{\cal I}_t- {\cal L }_t
 \ee
 where $t$ denotes the day number. Focusing our attention on the
long-term profit, we now evaluate the average long-term capital of
the company for a period of ${\cal N}$ working days. From the
above recursive relation, one simply finds by iteration:
 \be
 {\cal C}_{{\cal N}}={\cal C}_0 + p\sum_{t=1}^{{\cal N}}{\cal I}_t-
\sum_{t=1}^{{\cal N}}{\cal L}_t
\ee
 which determines the long-term capital in terms of aggregate loss and premium income.
 In order to predict the capital, we now average the above relation over many virtual
 realisations of future each of which corresponds to
${\cal N}$ working days. Let us first define such predictive
averaging process (hereafter referred to averaging ) for a generic
variable $\psi_t$ which for instance could be the number of
accidents or the number of issued policies in $t$-th day. We
define the average of ${\cal \psi}_t$ over $M$ virtual runs as
follows:
 \be
 < \psi_t>:=\frac{1}{M}[ \psi_t^{(1)} + \cdots + \psi_t^{(M)}]
\ee
 Where each $\psi_t^{(i)}$ is a random number drawn from a
specified distribution function. It can easily be verified that
for the quantities ${\cal I}_t$ and $ {\cal A}_t$ the averaging
does not depend on day number $t$. To see this explicitly let us
evaluate the average of the number of issued policies in the
$t$-th day. According to the above definition we have:
 \be
  <{\cal I}_t>= \frac{1}{M}[ {\cal I}_t^{(1)} + \cdots + {\cal I}_t^{(M)}]
 \ee

The term in the bracket is the sum of $M$ realisations of the
stochastic variable ${\cal I}_t$ and provided $M$ is large enough,
the sum simply converges to the average of the distribution
function which is ${\bar {\cal I}}$ for all days. Concerning this
fact, we now average over the long-term capital and obtain the
following relation where we have dropped the initial capital
${\cal C}_0$ for simplicity.
 \be
  <{\cal C}_N>= {\cal N}(p<{\cal I}>- <{\cal L}>)
 \ee
We define the critical value of premium $p_c$ at which the average
long-term profit equals zero $<{\cal C}_ {{\cal N}}>=0$. It is
worth obtaining $<{\cal L}>_t$ in terms of statistics of loss
events. The loss amount in the $t$-th day is an extended random
variable. By "extended" we mean that there are two sources of
randomness. Firstly the number of loss events and secondly the
loss amount for each loss event. In the appendix, it is shown, in
details, that the average loss amount for each day is simply the
average loss amount, in each single loss event, multiplied by the
average number of daily loss events i.e., $<{\cal L}_t>=\xi {\bar
{\cal A}}$. Concerning the above considerations one obtains the
following expression for the averaged long term capital : \be
 <{\cal C}>={\cal N}{\cal I}_0e^{-\frac{p-p_0}{\tau p_0} }(p-\theta\xi)
 \ee
 In our study, the numerical values of insurance parameters have
been set from the empirical data taken from  {\it Iranian
Insurance Industry} for the year 2000. Specifically, we analysed
the data taken from {\it Iran Insurance Company} the largest
company operating in Iran. The empirical data were: 1,819,935
insurance policy issued, 348961 loss events leading to 739.7
billion Rls loss. The premium amount was 378000 Rls on average.
Based on the above data, we set our parameter as: $\theta=0.19$,
${\xi= 2.12}$ million Rls and ${\cal I}_0=6070$. Also we take
${\cal N}=300$ active days. The following graph shows the
behaviour of $<{\cal C}>$ for different values of $\tau$.

\begin{figure}\label{Fig1}
\epsfxsize=8.5truecm \centerline{\epsfbox{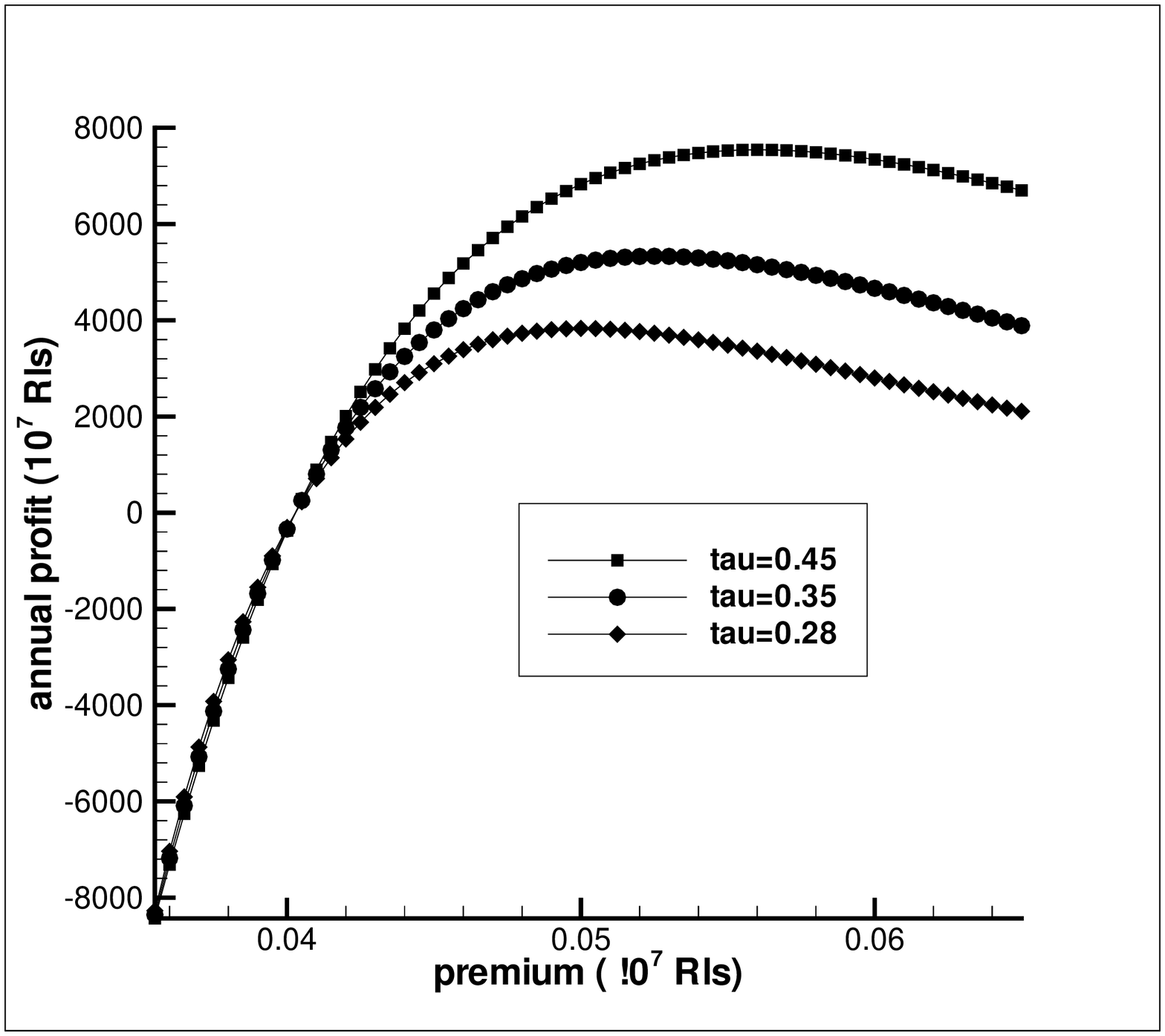}}
\end{figure}
\vspace{0.02 cm} {\small{Fig.~1: Averaged long term profit of the
virtual company for three values of $\tau$: 0.45, 0.35 and 0.28
which correspondingly refer to 20, 25 and 30 percent fall in the
number of issued policies upon increasing the premium by 10 percent.} }\\

Accordingly, two distinct regimes are identified: profit and loss.
There is crossover premium $p_c=\xi\theta$ above which the company
profits. If the premium is below $p_c$, the company is loss-making
and in the vicinity of the crossover point, the company is
profitless. The above graph gives us another useful knowledge. It
determines the optimal premium $p_{opt}$ at which the company's
capital is maximized. However, we should note that the
aforementioned conclusions are based on mean field approach and in
principle the company can not rely on these average-based
arguments. They need a quantified insight into the risk of premium
variations. A simple yet practical quantity which can give us an
approximate measurement of the risk is the variance of long term
profit in the predictive averaging process \cite{otovianni}.
 At this stage, we try to evaluate the variance of the long-term
capital ${\cal C}= <({\cal C}-{\bar {\cal C} })^2>$ where for
simplicity we have dropped the index ${\cal N}$ from the capital
index. After some straightforward mathematics we reach to the
following formula:
 $$
<({\cal C}-{\bar {\cal C} })^2>=p^2\sum_{t,t'=1}^{{\cal N}}<{\cal
I}_{t}{\cal I}_{t'}>-p^2{\cal N}^2{\bar {\cal I}}^2+  2p{\cal
N}^2{\bar {\cal I}}^2 \xi \theta $$ \be -2p\sum_{t,t'=1}^{{\cal
N}}<{\cal I}_{t}{\cal L}_{t'}>+\sum_{t,t'=1}^{{\cal N}}<{\cal
L}_{t}{\cal L}_{t'}>-{\cal N}^2{\bar {\cal I}}^2\xi^2\theta^2
 \ee

 It should be noted that even in the case where our stochastic
 variables have simple distribution functions, it is not yet an
 easy task to evaluate the correlation functions in the above
 formula. Recall the definition of $<{\cal I}_{t}{\cal I}_{t'}>$
 one simply finds:
 \be
<{\cal I}_{t}{\cal I}_{t'}>:=\frac{1}{M}[ {\cal I}_{t}^{(1)}{\cal
I}_{t'}^{(1)} + \cdots + {\cal I}_{t}^{(M)}{\cal I}_{t'}^{(M)}]
 \ee
which obviously differs from $<{\cal I}_{t}><{\cal I}_{t'}>$. To
proceed further, we separate the terms $t=t'$ in the above
relation and use the mean-field approximation in the remaining
terms $t\neq t'$ to substitute all the terms such as $<{\cal
I}_{t}{\cal I}_{t'}>$ by $<{\cal I}_{t}><{\cal I}_{t'}>$. We thus
obtain the following relation for variance of long term capital.
$$
 <({\cal C}-{\bar {\cal C} })^2>={\cal N}[p^2(<{\cal I}^2>-{\bar I}^2)-2p
 (<{\cal I}{\cal L}> -{\bar I}^2\xi\theta ) + $$
 \be
 <{\cal L}^2> -{\bar I}^2\xi^2\theta^2]
 \ee
The stochastic variable ${\cal I}$ has a uniform distribution
function therefore its variance is simply obtained via the
relation $\sigma^2=\frac{1}{12}(b-a)^2$ where $[a,b]$ denotes the
interval over which the variable is distributed. We recall that
$b-a={\bar {\cal I}}$ for ${\cal I}$. One can simply show that the
variance of daily loss is $\theta$ times the variance of ${\cal
I}$. Recalling the definition of the covariance between two
arbitrary stochastic variables $X$ and $Y$: \be
Cov(X,Y)=\frac{<XY>-<X><Y>}{\sigma_X\sigma_Y}
\ee
 It can be shown that inequality $-1 \leq Cov(X,Y)\leq 1$ holds.
 We thus write the remaining term in
the above equation as $\omega \sigma_{{\cal I}}\sigma_{{\cal L}}$
where $-1 \leq \omega \leq 1 $ denoted the covariance between
${\cal I}$ and ${\cal L}$. Putting everything together we simply
arrive at the following approximate equation for the variance of
long term capital: \be <({\cal C}-{\bar {\cal C} })^2>=\frac{{\cal
N}{\bar I}^2}{12}[p^2-2p\omega\xi\theta +\xi^2\theta^2] \ee It is
seen that risk value has a quadratic dependence on premium $p$ and
is proportional to the number of working days ${\cal N}$.

 The following graph depicts the behaviour of standard deviation as a
function of premium for some values of $\tau$:

\begin{figure}\label{Fig2}
\epsfxsize=8.5truecm \centerline{\epsfbox{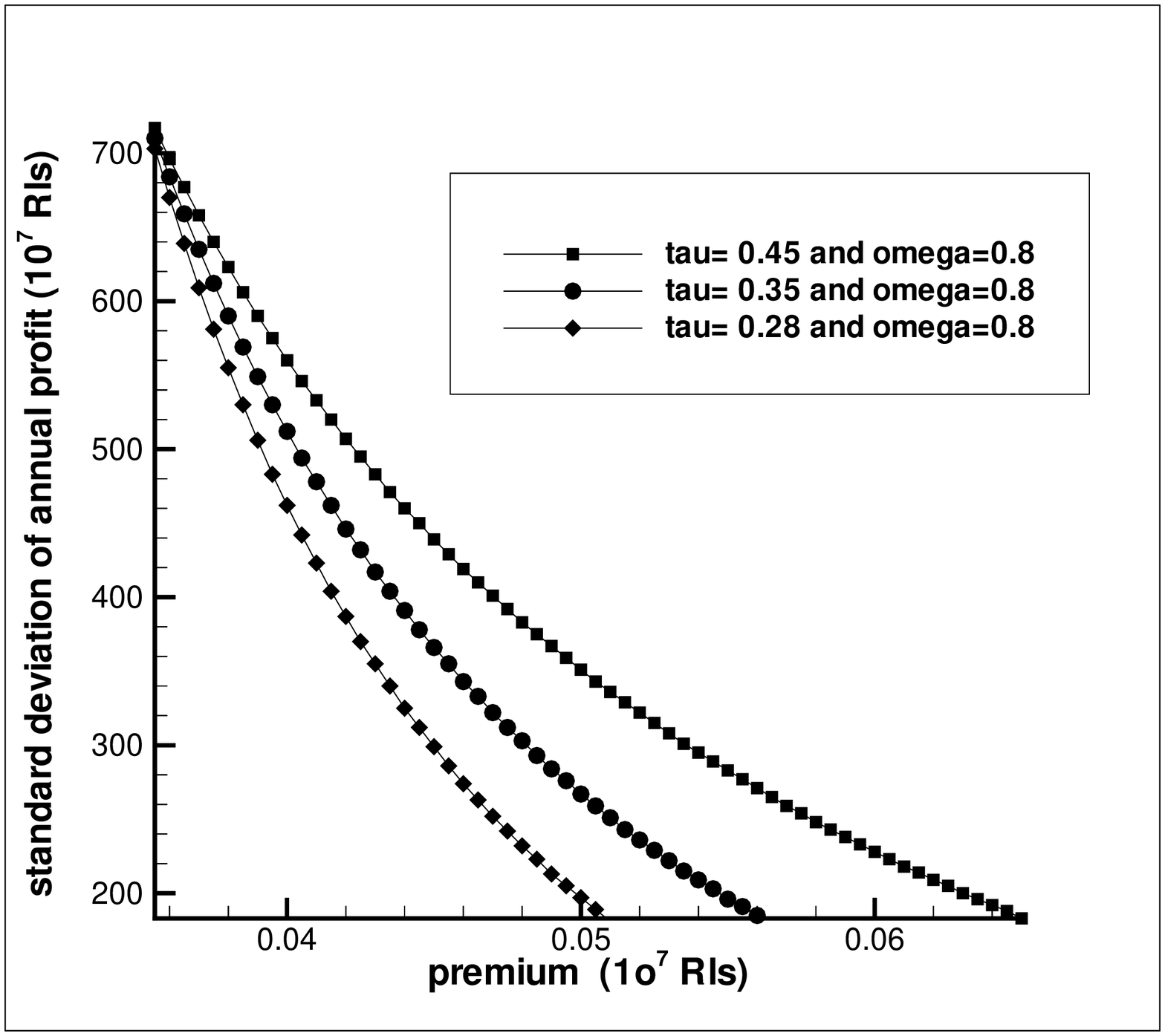}}
\end{figure}
\vspace{0.02 cm} {\small{Fig.~2: mean-field standard deviation of
annual capital for same three values of $\tau$ as in fig.1 .
$N=300$,
$\xi=2.12$ million Rls. and $\omega=0.8 .$} }\\

It is observed that risk value is an increasing function of
$\tau$. This indicates that for larger values of $\tau$ which
correspond to sharper decrease of insured number, we have a
less-valued variance in long term capital. This means that company
makes less profit but with more certainty, a situation which is
desirable for cautious managers who avoid risky decisions. It
would also be advantageous to look at the dependence of capital
variance on loss size average $\xi$. We recall that there are two
principal sources of uncertainty in the in-flow/out-flow of the
company's income. The first one is related to the predictive
number of insured, approximated by the parameter $\tau$ in our
model, which is the main in-flow portion of the company's capital.
The second one is incorporated with out-flow portion which is
dominated by loss size average $\xi$. The following graph exhibits
the long term capital variance for various values of $\xi$.

\begin{figure}\label{Fig3}
\epsfxsize=8.5truecm \centerline{\epsfbox{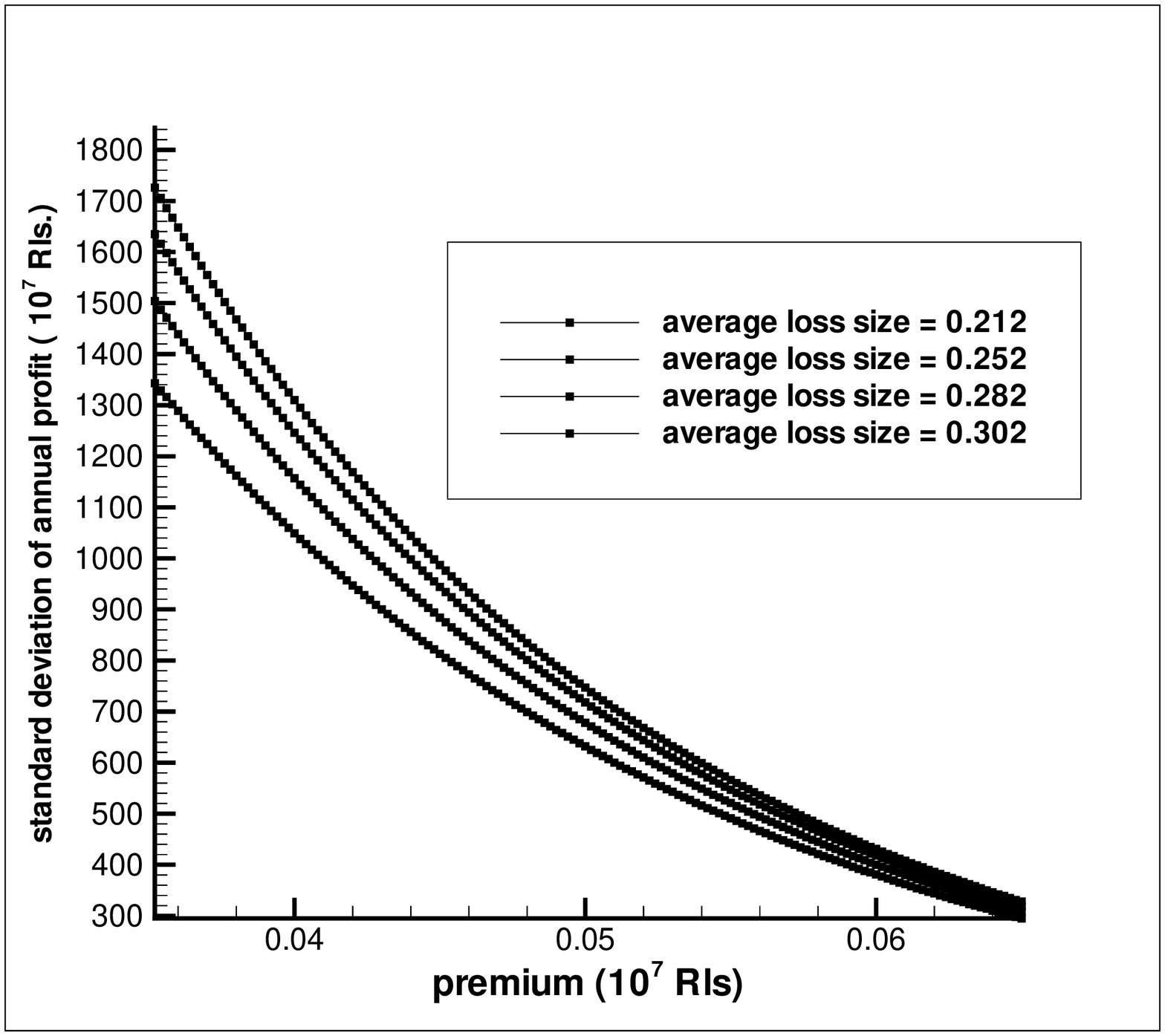}}
\end{figure}
\vspace{0.02 cm} {\small{Fig.~3: mean-field standard deviation of
annual capital for some values of $\xi$. $N=300$ and $\omega=0.3$ and $\tau=0.45$ .} }\\

The next concept we deal with is the bankruptcy. A useful quantity
in risk management is the company's ruin probability. This can be
expressed as the probability that the long term capital falls
below zero i.e., $P({\cal C}<0)$. To estimate the ruin
probability, we use of the so-called Tchebycheff inequality
\cite{reichel}. The Tchebycheff inequality establishes a relation
between the variance and the probability that a stochastic
variable, with finite average and variance, can deviate by an
arbitrary amount $\epsilon$ ($\epsilon
>0$) from the mean value: \be
 P(|z-<Z>|\geq \epsilon) \leq \frac{\sigma_{Z}^2}{\epsilon^2}
\ee where $\sigma_{Z}$ denotes the standard deviation of the
stochastic variable $Z$. Now let us estimate the probability that
for a prescribed $p$, the long term profit equals $-\lambda {\bar
{\cal C}}(p)$ where the dimension-less parameter $\lambda >0$
denotes the depth of ruin. According to Tchebycheff inequality if
we take $Z$ as ${\cal C}$ one simply finds: \be P({\cal
C}=-\lambda {\bar {\cal C} }) \precsim \frac{1}{2}P(|{\cal
C}-{\bar {\cal C} }|\geq {\bar {\cal C}}(1+\lambda))=
\frac{\sigma^2}{2 {\bar {\cal C}}^2(1+\lambda)^2}
 \ee
 Using equations
(9) and (6) for $\sigma$ and ${\bar {\cal C}}$ one obtains the
upper limit of ruin probability in terms of $p$ and $\lambda$. \be
 P({\cal C}=-\lambda {\bar {\cal C} }) \precsim \frac{p^2-2p\omega\theta\xi+\theta^2\xi^2}
 {24{\cal N}(p-\xi\theta)^2(1+\lambda)^2}
\ee
 The following graph exhibits the ruin probability
dependence on premium for various ruin depths.
\begin{figure}\label{Fig4}
\epsfxsize=8.5truecm \centerline{\epsfbox{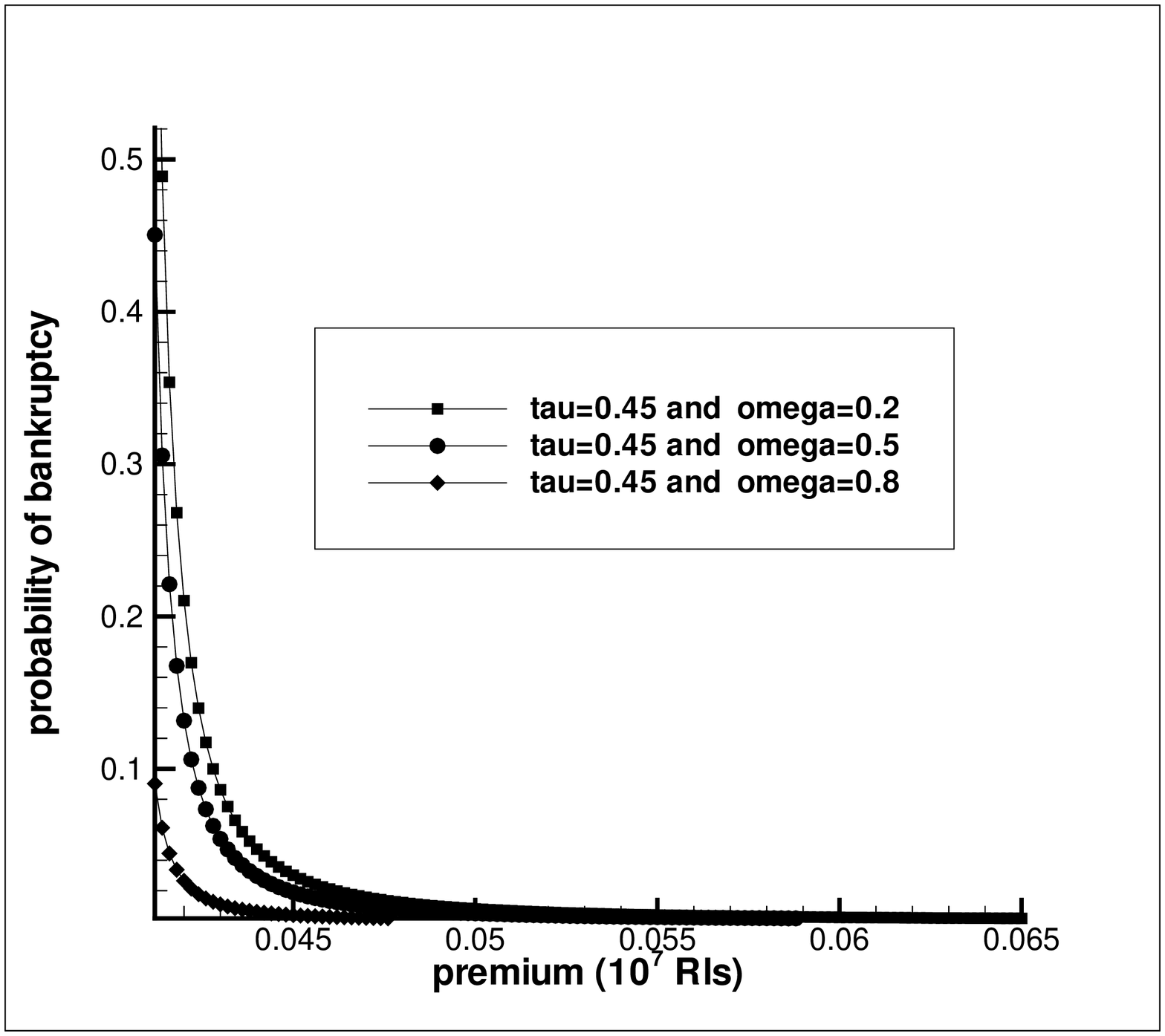}}
\end{figure}
\vspace{0.02 cm} {\small{Fig.~4: estimated ruin probability
according to Tchebycheff inequality. $\lambda$ determines the ruin
depth. Here $\lambda=0.1$. The rest of the parameter
are remained unchanged as in fig. 2} }\\


In order to find a deeper insight into the stochastic nature of
the problem, we have carried out simulations to obtain the
variance of ${\cal C}$. We have simulated the company's
performance for a year and obtain the annual income ${\cal
C}({\cal N})$. Figure five exhibits the simulated variance of
annual capital as a function of premium.

\begin{figure}\label{Fig5}
\epsfxsize=8.5truecm \centerline{\epsfbox{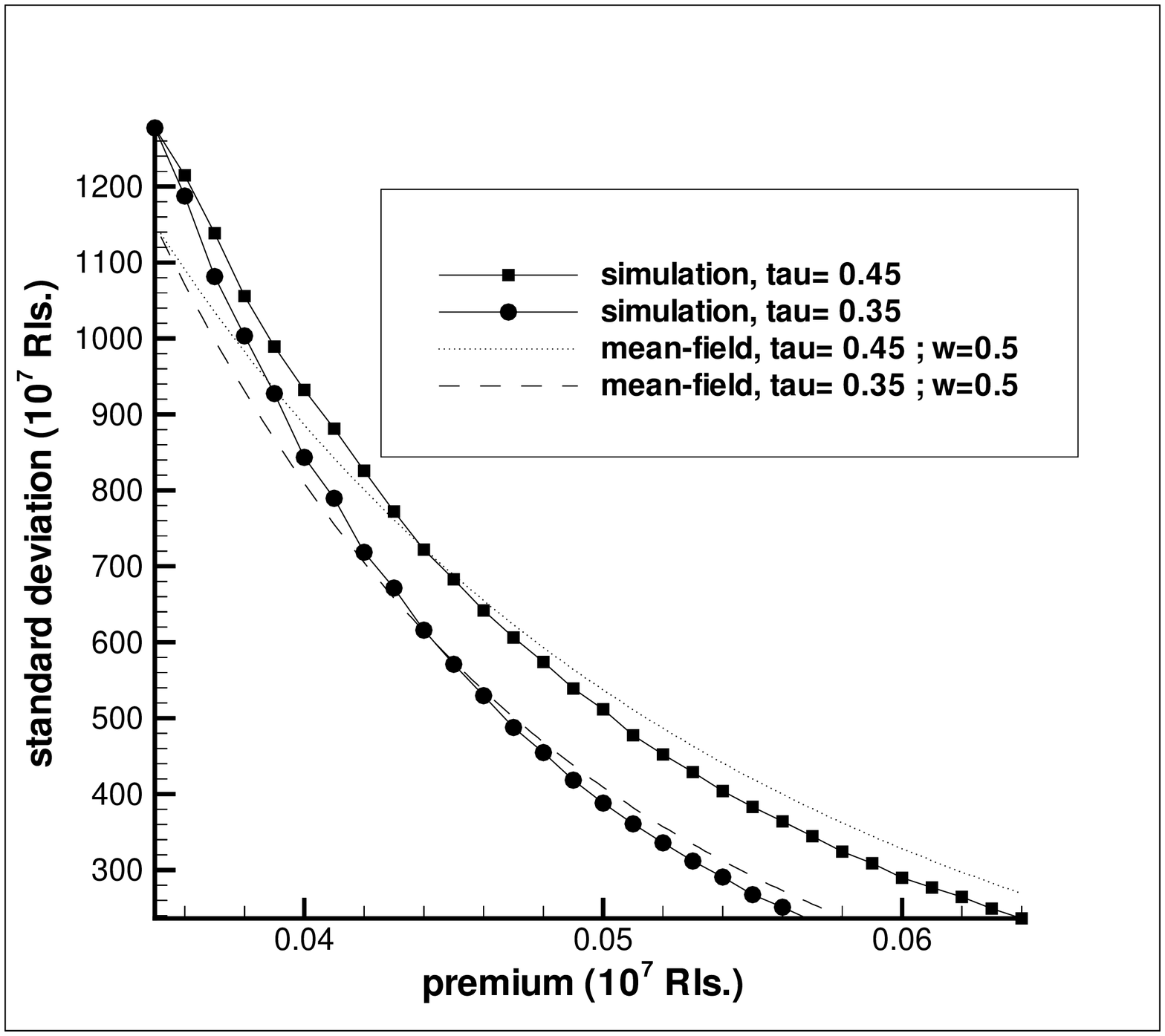}}
\end{figure}
\vspace{0.02 cm} {\small{Fig.~5: simulated standard deviation of
annual capital for the same values of $\tau$.
$N=300$, $\xi=2.12$ million Rls. } }\\

The simulation results show a notable deviation from those of
mean-field approach and gives $\omega\approx0.5$ as the best
fitting value for the covariance parameter $\omega$.

\section{Summary and Concluding Remarks}

In this paper, we have tried to present , in some detail, a set of
statistical facts emerging from the study of a virtual insurance
company. The financial operations of an insurer can be viewed in
terms of of a series of cash inflows and outflows. Premiums and
income from investments, together with certain other income, are
added to the reservoir of the assets, while the reservoir is
depleted by claim payments, expenses of running the business,
taxes and possibly other items of outflow. From a practical point
of view, it is of prime importance for every insurance company to
have a quantified measurement of the risk of insuring . In this
work, we have modelled the performance of a an insurance company
both analytically and numerically. Our results illustrates the
dependence of long-term profit in terms of insurance parameters
especially the premium. We have managed to give a quantified
estimation for the risk of premium increasing.


Finally we should point out some issues we have not discussed
here. The occurrence of loss events and its statistical features
is the prominent factor affecting the company's long term profit.
Throughout the paper we have assumed that the loss size
distribution obeys a simple statistics i.e. an exponential
distribution function $e^{-\frac{x}{\xi}}$. This is evidently a
localized distribution function with finite mean and variance
$\xi$ and $\xi^2$ respectively. However, the main source of ruin
in insurance industry is the occurrence of catastrophic events
corresponding to overwhelming loss sizes. Obviously ordinary
distribution functions such as Erlang fail to model these extreme
events. Investigation on this issue needs more exploration.

\section{ Acknowledgments}

We wish to express our gratitude to {\it Iranian Centre of
Actuarial Research } for 
providing us with the empirical data. We highly acknowledge
enlightening discussions with G.M. Sch\"{u}tz. We are also
thankful to K. Rahnama and A. Daghighi for fruitful discussions.

\bibliographystyle{unsrt}

\section {\bf Appendix}

In the section we obtain the relation $<{\cal L}_t>=\xi {\bar
{\cal A} }$. In order to numerically calculate the average of loss
amount in $t$-th day, we draw $l$ random numbers $n_t^{(1)},
\cdots, n_t^{(l)}$ from a uniform distribution function where
$n_t^{(i)}$ denotes the number of car accidents in the $i$-th
virtual realisation of $t$-th day. Let $z_{t,i}^{(j)}$ denote the
loss amount due to the $i$-th accident in the $j$-th virtual
realisation of $t$-th day. Accordingly we have :
 \be
 <{\cal L}>=\frac{z_1^{(1)}+\cdots+z_{n_1}^{(1)}+\cdots+z_1^{(l)}+\cdots+z_{n_l}^{(l)}}{l}
 \ee

where for simplicity we have dropped the day index $t$. The above
relation can be rearranged as:
 \be
\frac{n^{(1)}}{l}\frac{z_1^{(1)}+\cdots+z_{n_1}^{(1)}}{n^{(1)}}+\cdots+
\frac{n^{(l)}}{l}\frac{z_1^{(l)}+\cdots+z_{n_l}^{(l)}}{n^{(l)}}
\ee

Concerning the fact that each $n_{i}$ is considerably large, each
sum converges to $<Z>=\xi$ therefore we have:
 \be
  <{\cal L}>=\xi[\frac{n^{(1)}+\cdots+n^{(l)}}{l}]
 \ee

Now the term in the bracket is simply the average of ${\cal A}$
therefore we have the relation $<{\cal L}>=\xi{\bar {\cal A}}$

\end{multicols}{2}
\end{document}